\documentstyle[preprint,aps,tighten,epsf]{revtex}

\begin{document}

\draft

\title{Coulomb blockade peak spacing fluctuations in deformable
quantum dots: a further test to Random Matrix Theory}

\author{Ra\'ul O. Vallejos,$^1$ Caio H. Lewenkopf,$^1$ 
	and Eduardo R. Mucciolo$^2$}

\address{$^1$
	 Instituto de F\'{\i}sica, 
	 Universidade do Estado do Rio de Janeiro, \\
	 R. S\~ao Francisco Xavier, 524, CEP 20559-900 Rio de Janeiro, 
	 Brazil\\
	 $^2$
	 Departamento de F\'{\i}sica, 
	 Pontif\'{\i}cia Universidade Cat\'olica do Rio de Janeiro,\\
	 Cx.P. 38071, 22452-970 Rio de Janeiro, Brazil}

\date{\today}

\maketitle

\begin{abstract}
We propose a mechanism to explain the fluctuations of the ground state
energy in quantum dots in the Coulomb blockade regime. Employing the
random matrix theory we show that shape deformations may change the
adjacent peak spacing distribution from Wigner-Dyson to nearly
Gaussian even in the absence of strong charging energy
fluctuations. We find that this distribution is solely determined by
the average number of anti-crossings between consecutive conductance
peaks and the presence or absence of a magnetic field. Our mechanism
is tested in a dynamical model whose underlying classical dynamics is
chaotic. Our results are in good agreement with recent experiments and
apply to quantum dots with spin resolved or spin degenerate states.
\end{abstract}

\pacs{PACS numbers: 73.23.Hk, 05.45.+b}



\narrowtext

Most phenomena in mesoscopic electronic transport are well understood
invoking the picture of a single quantum particle moving coherently in
a disordered or complex potential. In the case of semiconductor
quantum dots, the complexity of the background or confining potential
is manifest in the chaotic nature of the underlying electronic
classical dynamics. A large amount of literature \cite{LesHouches}
indicates that the statistical properties of the spectra of chaotic
systems are successfully modeled by random matrix theory (RMT)
\cite{Mehta91}. When applied to transport in open ballistic quantum
dots of irregular shape, this theory predicts a strong universal
behavior for the statistical fluctuations of the conductance, similar
to that found in disordered systems \cite{Baranger}. Over the last
years, several experiments confirmed this behavior, making open
electronic cavities one of the paradigms of quantum chaos
\cite{Beenakker97,Guhr97}.

Recent experiments \cite{Chang96,Folk96} showed that RMT is also very
successful in predicting the distribution of Coulomb blockade peaks
heights in nearly isolated, many-electron quantum dots at low
temperatures and zero bias \cite{Jalabert92}. In this case, the
theoretical description is based on the so-called `constant
interaction' (CI) model \cite{Kastner92}, where the electron-electron
interaction is taken into account through a fixed (capacitive)
charging energy term in the Hamiltonian. The electronic many-body wave
function is then formed by filling single-particle eigenstates whose
fluctuations reflect the chaotic nature of the classical motion within
the dot.

However, new data related to the many-body ground state energy
fluctuations in these systems are at variance with the standard RMT
predictions. It was observed that adjacent peak spacing energies do
not fluctuate according to the celebrated Wigner surmise
\cite{Sivan96,Simmel97,Patel97}. Moreover, the existence of strong
correlation among the heights of nearby peaks and the apparent lack of
spin degeneracy seemed to indicate that other effects, such as wave
function scarring \cite{Stopa97}, interaction-induced correlations,
and charging energy fluctuations \cite{Sivan96,Blanter97,Berkovits97},
also play important roles.

The aim of this letter is to show that, although effects beyond the
single-particle approach may be relevant to the general understanding
of quantum dots, there is yet a simple mechanism based on the CI model
and RMT which explains the basic feature found in
Refs.~\onlinecite{Sivan96,Simmel97,Patel97}, namely, the approximately
Gaussian form of the peak spacing distribution. The essential point
which has been missing in all previous analyses of the problem is that
the gate voltage, swept to produce a series of Coulomb blockade peaks,
also continuously deforms the confining potential seen by the
electrons in the quantum dot \cite{Hackenbroich97}. As a consequence,
ground state energies, which are measured from the positions of
consecutive peaks, correspond to {\it different} parametric
realizations of the one-body Hamiltonian. Since adjacent energy
eigenvalues tend to lose correlation rapidly as their parametric
distance increases \cite{Simons93}, the experiments in question are
likely to be measuring the spacing distribution of {\it uncorrelated}
chaotic eigenstates. If the dot shape dependence on the gate voltage
is sufficiently strong, we find that the peak spacing distribution
approaches a Gaussian, regardless of the existence or not of spin
degeneracy in the single-particle spectrum.
 
In order to present the quantitative aspects of our argument, we begin
indicating the condition for the occurrence of a conductance peak
within the CI model \cite{Beenakker91}:
\begin{equation}
\label{resonance}
E_n + n U - e \eta V_g = E_F,
\end{equation}
where $E_n$ is the $n^{\mbox{\scriptsize th}}$ single-particle energy
level ($n \gg 1$), $U$ is the dot charging energy, $V_g$ is the gate
potential, and $E_F$ is the Fermi energy (chemical potential) in the
leads. The coefficient $\eta$ is a function of the capacitance matrix
elements of the dot \cite{Glazman89}.

In all experimental setups where plunger gates are used to define both
shape and depth of the confining potential, one expects that
variations in the potential $V_g$ will continuously deform the
dot. This is the actual situation at least for the experiments in
Refs.~\onlinecite{Simmel97,Patel97}. To establish the direct relation
between $V_g$ and the dot shape we need to know the experimental set
up in detail. While such study is desirable, it is not crucial for our
semi-quantitative analysis and will not be pursued here. We therefore
simplify the problem parametrizing the shape deformations by a generic
variable $X$, such that $E_n=E_n(X)$.

In view of Eq.~(\ref{resonance}), the difference between the positions
(in gate voltage) of two consecutive peaks is
\begin{equation}
\delta V_g(n) = (e\eta)^{-1}[ U + E_n(X_n) - E_{n-1}(X_{n-1})].
\end{equation}
Subtracting the constant charging term, the single-particle
contribution to the energy spacing can be written as
\begin{eqnarray}
\label{key}
\delta E_g(n) & \equiv & e\eta\ \delta V_g(n) - U \nonumber \\ & = &
E_n (X_{n-1}) - E_{n-1}(X_{n-1}) + \delta E_n ,
\end{eqnarray}
with $\delta E_n = E_n(X_n) - E_n(X_{n-1})$. It is clear from
Eq.~(\ref{key}) that the fluctuations in $\delta E_g(n)$ {\it should
not} obey the Wigner-Dyson statistics when $\delta E_{n}$ also
fluctuates on a scale larger or of the order of the mean
single-particle level spacing $\Delta = \langle E_n(X_{n-1}) -
E_{n-1}(X_{n-1}) \rangle$. To understand this point, let us consider
that a gate voltage variation $\delta V_g$, sufficient to add one
electron to the dot, also deforms the confining potential by $\delta
X$. Three illustrative situations are depicted in the single-particle
spectrum shown in Fig. 1. The states indicated by the black dots
fulfill Eq.~(\ref{resonance}) for different values of $n$. The
sequence (a) corresponds to the filling of single-particle states when
$\delta X = 0$, in which case the level spacings should indeed follow
the Wigner-Dyson statistics. When $\delta X \ne 0$ and the deformation
is moderate, we may have instead a sequence like (b). Now each filled
state is slightly shifted with respect to its predecessor, level
repulsion is weakened ,and deviations from the Wigner-Dyson statistics
begin to appear. Finally, when deformations are strong, a sequence
like (c) may occur: Consecutive filled states fluctuate independently
and there is hardly any sign of level repulsion. In fact, in the
presence of an overall downward drift, level spacings can be
``negative''.

The quantity which determines the form of the distribution of $\delta
E_g$, $P(\delta E_g)$, is the ratio between $\delta X$ and the typical
distance between anti-crossings in parameter space, $X_c$. In what
follows we study $P(\delta E_g)$ as a function of $\delta X/X_c$.

Peak-to-peak fluctuations in $\delta X$ are of the order $\Delta/U$
and can be neglected when $U\gg\Delta$, which is the regime relevant
to the experiments. Thus, let us assume that $\delta X$ independs of
$n$. This allows us to calculate $P(\delta E_g)$ analytically in the
perturbative regime. For this purpose, we model the parametric
dependence of the electronic Hamiltonian as
\begin{equation}
H_{\mbox{\scriptsize tot}}(X +\delta X) = H + \delta X\ K,
\end{equation}
where $H$ and $K$ are large, independent $N\times N$ matrices
belonging to the proper Gaussian ensemble. The orthogonal ensemble
(GOE, $\beta=1$) is used when no magnetic field $B$ is present,
whereas for $B\neq 0$ we use the unitary ensemble (GUE, $\beta=2$).
We use first-order perturbation theory to describe the regime of small
$\delta X/X_c$ and write $\delta E_{n} \approx \delta X
K_{nn}$. Therefore,
\begin{equation}
\label{firstPE}
P(\delta E_g) \approx \Big\langle \delta \biglb( \delta E_g - \bigl(
E_n - E_{n-1} + \delta X K_{nn} \bigr) \bigrb) \Big\rangle_{H,K},
\end{equation}
where $\langle\cdots\rangle_{H,K}$ represents the ensemble averages
over $H$ and $K$. The average over $K$ yields a Gaussian factor that
depends on the first two moments of $K_{nn}$. We choose $\langle
K_{nn} \rangle = 0$ and $\langle (K_{nn})^2 \rangle =
\Delta^2/X_c^2$. The former prevents level drift, while the latter
corresponds to identifying the typical distance between anti-crossings
with the inverse root mean square derivative of the energy levels with
respect to deformations \cite{Simons93}, namely, $X_c = \Delta/\sqrt{
\langle (dE_n/dX)^2 \rangle}$. In these terms,
\begin{equation}
P(\delta \varepsilon_g) \approx \left\langle \frac{1}{\sqrt{2\pi} x}
\exp\left[ -\frac{(\delta \varepsilon_g - \varepsilon_n +
\varepsilon_{n-1})^2}{2 x^2} \right] \right\rangle_{H},
\label{interm}
\end{equation}
where $x \equiv \delta X/X_c$. For the sake of simplicity, we have
rescaled all energies in units of $\Delta$, introducing $\delta
\varepsilon_g \equiv \delta E_g/\Delta$ and $\varepsilon_n =
E_n/\Delta$ for all $n$. Since Eq.~(\ref{interm}) depends only on the
spacing between neighboring levels, the average over $H$ can be
replaced by the convolution
\begin{equation}
\label{quasela}
P(\delta \varepsilon_g) \approx \int_0^\infty ds P^{WD}_\beta(s)
\frac{1} {\sqrt{2\pi} x} \exp \left[ - \frac{(\delta \varepsilon_g -
s)^2}{2 x^2} \right].
\end{equation}
$P^{WD}_\beta(s)$ denotes the Wigner-Dyson nearest-neighbor spacing
distribution (WD), corresponding to the GOE (GUE) for $\beta =1$
($\beta=2$) \cite{Mehta91}. Notice that in the limit of $x \rightarrow
0$ the Gaussian factor becomes a $\delta$-function and $P(\delta
\varepsilon_g)$ equals the WD, as expected.

The analytical evaluation of Eq.~(\ref{quasela}) is possible when we
approximate the WD by the Wigner surmise $P^{WD}_\beta(s) \approx
a_\beta s^\beta e^{-b_\beta s^2}$, with $a_{1(2)} = \pi/2(32/\pi^2)$
and $b_{1(2)} = \pi/4 (4/\pi)$. For the GOE, introducing $\xi_1^2
\equiv \pi/4 + 1/(2x^2)$, the result is
\begin{eqnarray}
P_{GOE} (\delta \varepsilon_g) & \approx & \frac{\pi^{1/2}} {2^{5/2} x
\xi_1^2} \exp \left( -\frac{ \delta \varepsilon_g^2} {2x^2} \right)
\Bigg\{ 1 + \frac{\pi^{1/2} \delta \varepsilon_g} {2 x^2 \xi_1}
\nonumber \\ & & \times \exp \left( \frac{\delta
\varepsilon_g^2}{4x^4\xi_1^2} \right) \left[ 1 + \mbox{erf} \left(
\frac{\delta \varepsilon_g}{2 x^2 \xi_1} \right) \right] \Bigg\},
\label{PGOE}
\end{eqnarray}
whereas for the GUE, with $\xi_2^2 \equiv 4/\pi + 1/(2x^2)$, we have
\begin{eqnarray}
P_{GUE}(\delta \varepsilon_g) & \approx & \frac{2^{5/2} \delta
\varepsilon_g} {\pi^{5/2} x^3 \xi_2^4} \exp \left( -\frac{\delta
\varepsilon_g^2}{2x^2} \right) \Bigg\{ 1 + \frac{\pi^{1/2}
(2x^4\xi_2^2 + \delta \varepsilon_g^2)} {2 \delta \varepsilon_g \xi_2}
\nonumber \\ & & \times \exp \left( \frac{\delta \varepsilon_g^2}
{4x^4\xi_2^2} \right) \left[ 1 + \mbox{erf} \left( \frac{\delta
\varepsilon_g}{2 x^2 \xi_2} \right) \right] \Bigg\}.
\label{PGUE}
\end{eqnarray}
Notice that the Gaussian factor in Eq.~(\ref{quasela}) has the
unsatisfactory feature that it becomes indefinitely broad for
increasing values of $x$. This is so because the perturbation theory
cannot describe consecutive anti-crossings that stabilize the level
position and therefore is limited to $x\ll 1$.

We have also performed extensive numerical simulations of parametric
realizations of Gaussian ensembles of $N = 500$ random matrices to
obtain $P(\delta \varepsilon_g)$. For $x < 0.25$ the agreement with
the analytical results is very good, with deviations becoming larger
with increasing $x$, as expected. Moreover, only for $x < 0.25$ we
have found that variations of $x$ change significantly the shape of
$P(\delta \varepsilon_g)$. The numerical results indicate that, for
$x>1$, $P(\delta \varepsilon_g)$ approaches a Gaussian distribution
whose variance saturates slowly as $x$ increases, as can be observed
in Fig.~2. Thus, the utility of the analytical results is that they
capture the essence of the change in shape of $P(\delta
\varepsilon_g)$, ceasing to be quantitative for large values of $x$.

The reason why $P(\delta \varepsilon_g)$ tends to a Gaussian may be
explained in the following way. First, for a sufficiently strong
deformation, we expect $E_n (X_n)$ and $E_{n-1} (X_{n-1})$ to become
independent. Second, we have empirically found from our simulations
that the position of each level taken alone fluctuates according to a
Gaussian distribution, with a certain ensemble-dependent standard
deviation $\sigma_\beta \sim O(1)$. In view of Eq.~(\ref{key}), we
conclude that
\begin{equation}
P(\delta \varepsilon_g) \Big|_{x\rightarrow \infty} \longrightarrow
\frac{1}{\sqrt{4\pi} \sigma_\beta} \exp \left[ -\frac{ (\delta
\varepsilon_g - 1)^2} {4 \sigma_\beta^2} \right].
\end{equation}
(The variance of the unitary case is always smaller than the
orthogonal one because the level repulsion grows with $\beta$.) We
remark that some analytical work \cite{Forrester97}, as well as the
Dyson Brownian motion model \cite{Mehta91} hint the idea that the
position of each level taken alone is Gaussian distributed, although
we found no rigorous proof in the literature.

So far we have not considered the possibility of having spin
degenerate eigenstates. In other words, the discussion focused on the
case that, due to some strong exchange contribution, the mean-field
solution of the many-body problem led to states with broken spin
degeneracy. Let us discuss now the opposite case of spin degenerate
levels, when the nearest-neighbor level spacing distribution is
\begin{equation}
P^{SD}_\beta(s) = \frac{1}{2}\delta(s) + \frac{1}{2}P^{WD}_\beta(s).
\end{equation}
For $\delta X=0$ this leads to a bimodal sequence of peaks spacings
(and also of peak heights). However, in analogy to the previous
discussion, as $\delta X$ increases, even spin degenerate peaks will
correspond to different dot shapes. Consequently, the bimodal
structure will be rapidly destroyed. As $\delta X$ approaches the
scale of one avoided crossing, $P(\delta \varepsilon_g)$ will tend to
a Gaussian. For $x>1$ the only significant difference between the case
of broken and preserved spin degeneracy is the variance of the
distributions with respect to $\Delta$, as shown in Fig. 2.

We test our mechanism by modeling the dynamics of an electron in a
quantum dot using the conformal billiard \cite{Berry86}. The shape of
the billiard is determined by the image of the circle of unit radius
in the complex plane $z$ under the conformal mapping $w(z)=(z+bz^2+c
e^{i\delta}z^3)/(1 + 2b^2 + 3c^3)^{1/2}$. We studied the case $b=0.2$
and $c=0.2$, sweeping the parameter $\delta$ in $[\pi/2,\pi]$. For
this parameter range the conformal billiard is known to display
chaotic motion in its classical limit \cite{Bruus94}. Another
convenient feature of this model is the readiness to compute
eigenvalues and eigenfunctions when the billiard is threaded by an
Aharonov-Bohm flux of arbitrary strength. This allows us to study the
cases of preserved and broken time-reversal symmetry with equal
numerical effort. We have considered eigenvalues ordered in ascending
energy ranging from the 200$^{\mbox{\scriptsize th}}$ up to the
250$^{\mbox{\scriptsize th}}$. In this energy window, we followed the
levels in parameter space to construct our histograms, observing that
$\Delta$ and $X_c$ stay almost constant over the whole
energy-parameter window \cite{Bruus97}. The scaled results are in good
agreement with our theoretical predictions based on RMT. In Fig. 3 we
illustrate this statement showing the case of spin resolved
degeneracy.

Three experiments \cite{Sivan96,Simmel97,Patel97} have obtained peak
spacing distributions that fit Gaussian curves rather well. Our
results are consistent with these findings if we take $x>1$,
indicating that in the experimental setups there is at least one
single-particle anti-crossing when the gate voltage is swept between
peaks. Some difficulties arise when we try a more precise quantitative
comparison. In particular, in Ref.~\cite{Patel97} the authors found
weak upward deviations from the Gaussian behavior at {\it both} tails
of the distribution which cannot be reproduced within RMT and are not
seen in our simulations. While the spectral statistics of the
conformal billiard is rather universal, it is customary for chaotic
systems to exhibit strong scarring and other non-universal features,
resulting in levels that fluctuate weakly with shape deformations and
do not follow RMT statistics. Although such levels are few, they may
lead to deviations in the tails of $P(\delta\varepsilon_g)$. This
subject is currently under investigation. Thermal smearing was also
not taken into account in our analysis and may lead to a small
distortion of the peak spacing distribution.

In conclusion, we have shown that the statistics of Coulomb blockade
peak spacing distribution in chaotic quantum dots depends on shape
deformations induced by variations in the gate voltage. Even when no
fluctuations in the charging energy are present, provided that
deformations are strong, the distribution of peak spacings is nearly
Gaussian and quite distinct from the Wigner surmise, in qualitative
accordance with experiments. We propose that the absence of
spin-degeneracy bimodal structures seen in the experiments may also be
caused by deformations. Although we have only considered the situation
of constant charging energy in this letter, we believe that
deformations should also be incorporated when wave function scarring
and many-body effects beyond mean-field are present.

We thank H. Bruus for providing us with the code for diagonalizing the
conformal billiard. We gratefully acknowledge the financial support of
FAPERJ, CNPq, and PRONEX (Brazil).



\begin{figure}
\setlength{\unitlength}{1mm}
\begin{picture}(160,160)(0,0)
\put(0,20){\epsfxsize=16cm\epsfbox{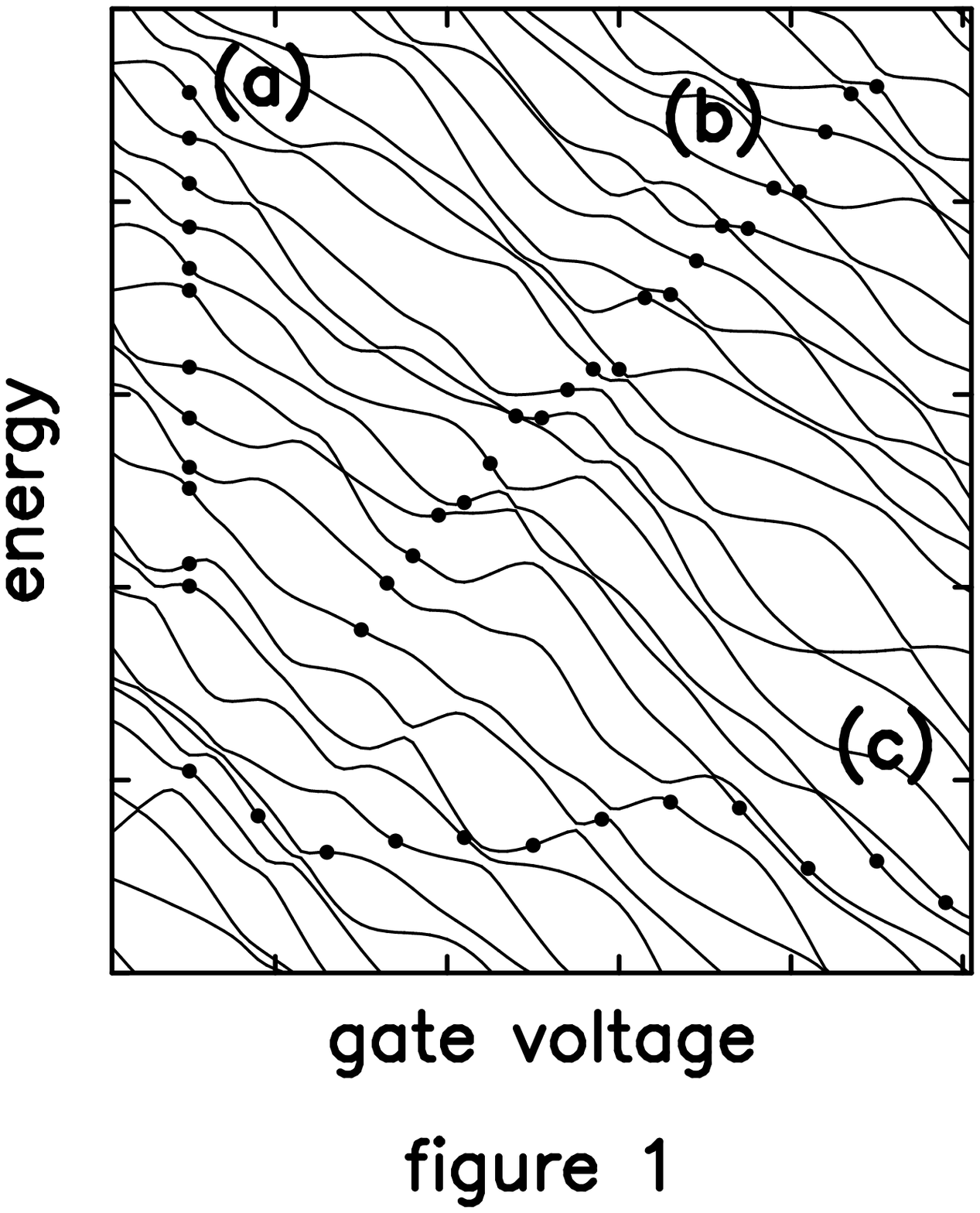}}
\end{picture}
\caption{Single-particle spectrum as a function of the gate voltage
(or dot shape deformation).}
\end{figure}

\newpage

\begin{figure}
\setlength{\unitlength}{1mm}
\begin{picture}(160,160)(0,0)
\put(0,30){\epsfxsize=16cm\epsfbox{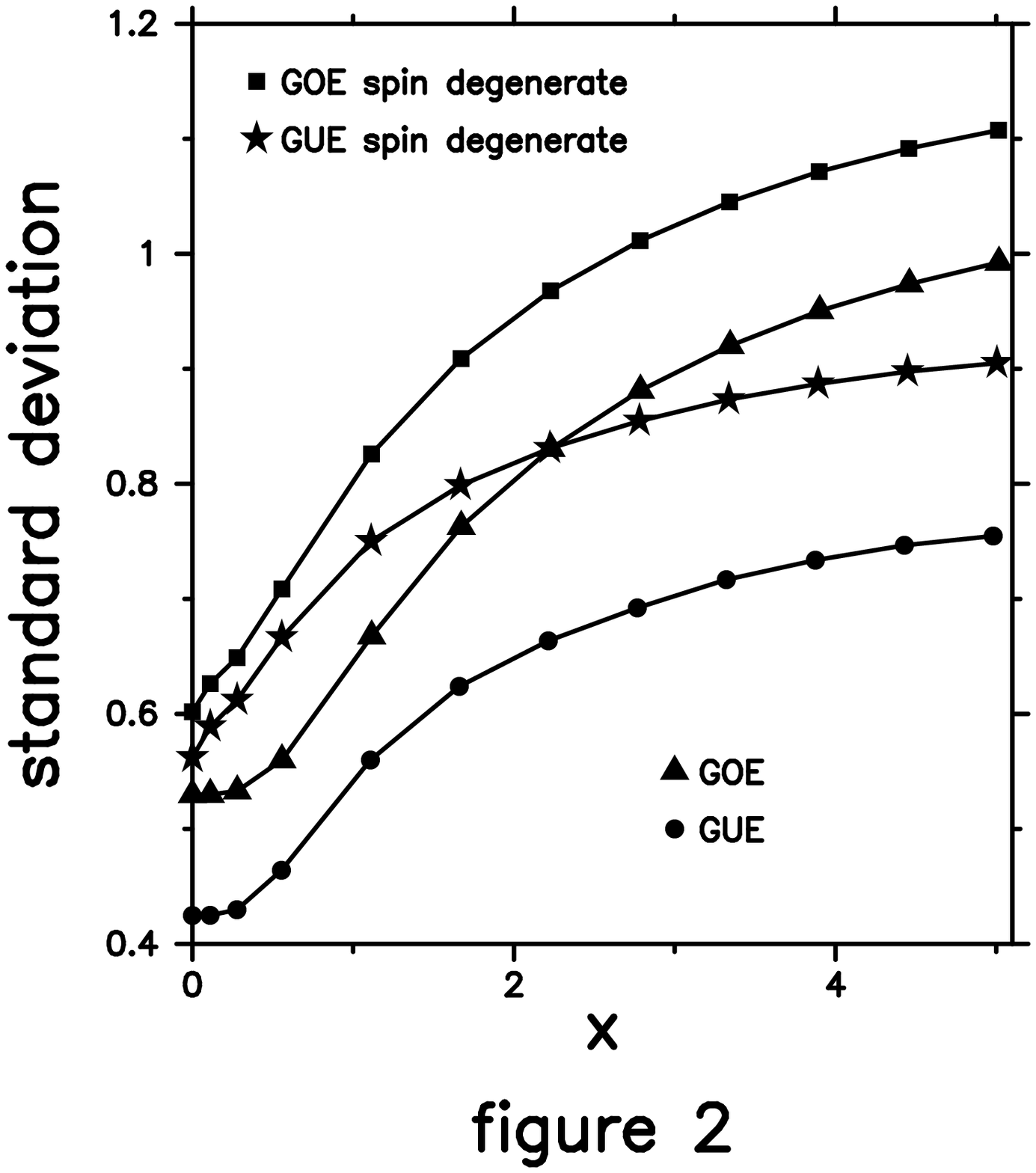}}
\end{picture}
\caption{Standard deviation of $P(\delta \varepsilon_g)$ as a function
of $x$. Squares correspond to GOE and stars to GUE for the spin
degenerate case. Triangles correspond to GOE and circles to GUE for
the spin resolved case.}
\end{figure}

\newpage

\begin{figure}
\setlength{\unitlength}{1mm}
\begin{picture}(160,160)(0,0)
\put(0,20){\epsfxsize=16cm\epsfbox{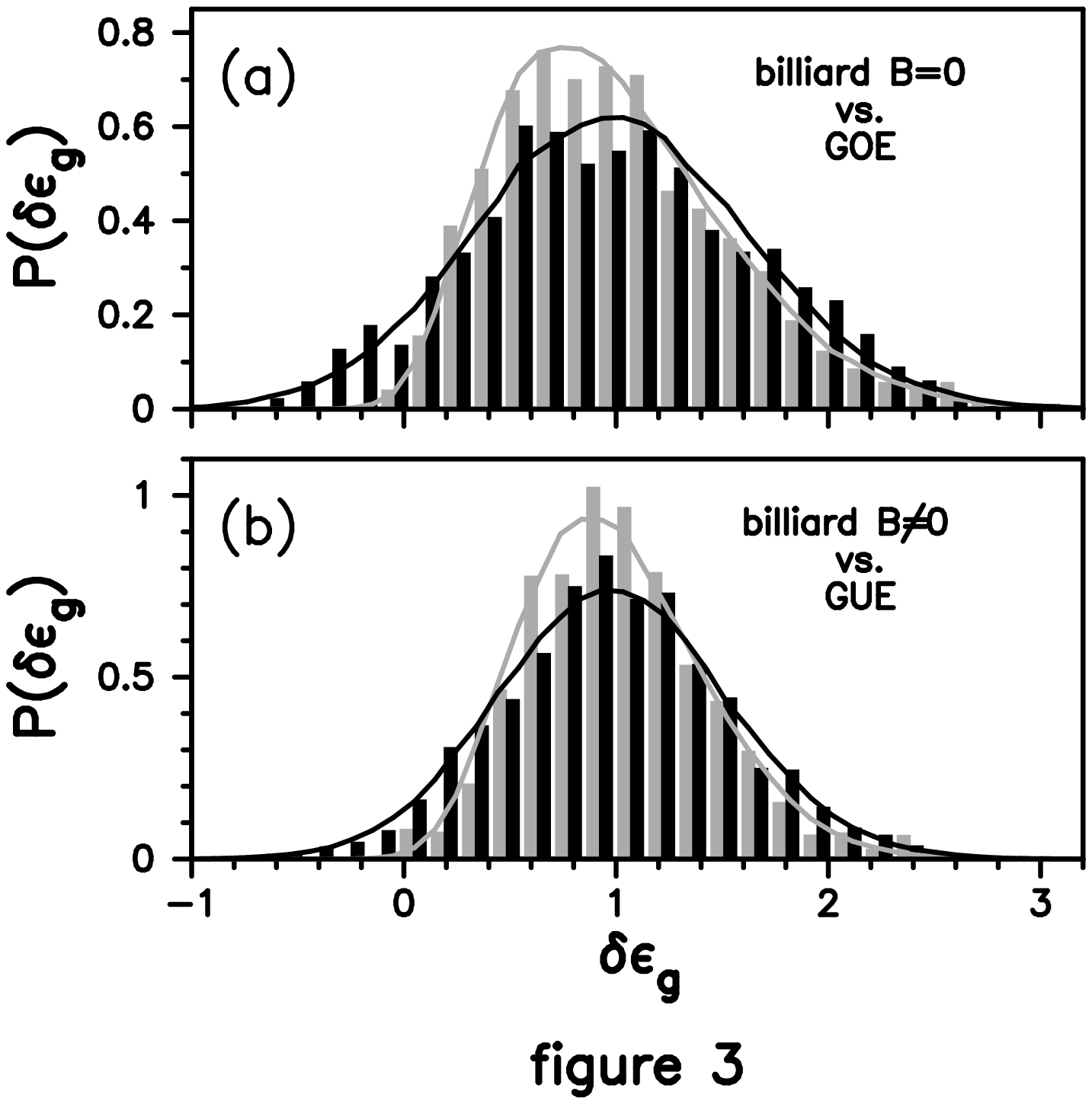}}
\end{picture}
\caption{Comparison between $P(\delta\varepsilon_g)$ obtained from the
conformal billiard (histograms) and our numerical simulations (solid
lines) for the spin resolved case: (a) GOE for $x=0.25$ (in gray) and
$x=1$ (in black); (b) the same for GUE.}
\end{figure}




\end{document}